\documentclass[12pt,
superscriptaddress,]{iopart}
\usepackage{iopams}  
\expandafter\let\csname equation*\endcsname=\relax 
\expandafter\let\csname endequation*\endcsname=\relax
\usepackage{amsmath}
\usepackage{bbm}
\usepackage{graphicx, epstopdf}
\usepackage{mathtools}
\usepackage{dcolumn}
\usepackage{bm}
\usepackage{hyperref}
\usepackage{cleveref} 
\usepackage{color}
\usepackage[numbers,sort&compress]{natbib} 
\newcommand{\revise}{}
\newcommand{\change}{}

\begin{document}

\title{Distribution and pressure of active L\'{e}vy swimmers under confinement}

\author{Tingtao Zhou$^1$, Zhiwei Peng$^2$, Mamikon Gulian$^3$, 
and John F. Brady$^{1,2}$}
\address{$^1$ Division of Engineering and Applied Science, California Institute of Technology, Pasadena, California 91125, USA}
\address{$^2$ Division of Chemistry and Chemical Engineering, California Institute of Technology, Pasadena, California 91125, USA}
\address{$^3$ Center for Computing Research, Sandia National Laboratories, Albuquerque, New Mexico 87123, USA}

\vspace{10pt}
\begin{indented}
\item[]Mar 2021
\end{indented}

\begin{abstract}
Many active matter systems are known to perform L\'{e}vy walks during migration or foraging. Such superdiffusive transport indicates long-range correlated dynamics. These behavior patterns have been  observed for microswimmers such as bacteria in microfluidic experiments, where Gaussian noise assumptions are insufficient to explain the data. We introduce \textit{active L\'evy swimmers} to model such behavior. \change{The focus is} on ideal swimmers that only interact with the walls but not with each other, which reduces to the classical L\'evy walk model  but now under confinement.
We study the density distribution in the channel and force exerted on the walls by the L\'evy swimmers, where the boundaries require proper explicit treatment. 
We analyze stronger confinement via a set of coupled kinetics equations and the swimmers’ stochastic trajectories.
Previous literature demonstrated that power-law scaling in a multiscale analysis in free space results in a fractional diffusion equation. We show that in a channel, in the weak confinement limit active L\'evy swimmers are governed by a modified Riesz fractional derivative. Leveraging recent results on fractional fluxes, we derive steady state solutions for the bulk density distribution of active L\'evy swimmers in a channel, 
and demonstrate that these solutions agree well with particle simulations. The profiles are non-uniform over the entire domain, in contrast to constant-in-the-bulk profiles of active Brownian and run-and-tumble particles. 
Our theory provides a mathematical framework for L\'evy walks under confinement with sliding no-flux boundary conditions and 
provides a foundation for studies
of interacting active L\'evy swimmers.

\end{abstract}

\vspace{2pc}

\submitto{\jpa}

\section{Introduction}
Active matter refers to systems of self-propelled particles or \textit{swimmers}, such as motile cells or synthetic Janus particles. In contrast to passive Brownian systems, active particles exhibit intriguing behaviors including accumulation at confining boundaries, upstream swimming in Poiseuille flow, and motility-induced phase separation \citep{Elgeti_2015,Bechinger2016RevModPhys, cates2015motility}. The non-equilibrium nature of active matter poses a number of theoretical challenges and has attracted considerable interest in the past few decades. To aid in the understanding and prediction of the dynamics of active matter, various theoretical models have been proposed, either at the particle scale \citep{Bechinger2016RevModPhys} or the macroscopic scale \citep{Marchetti2013,Julicher_2018}. Among them, stochastic dynamical models such as active Brownian particles (ABP) and run-and-tumble particles (RTP) are widely used \citep{Elgeti_2015, Bechinger2016RevModPhys, romanczuk2012active}. In these stochastic models, an active particle self-propels with an intrinsic speed $U_0$ and reorients on a timescale $\tau$ due to either continuous rotary diffusion (ABP) or discrete tumbling events (RTP). For ABPs and RTPs, the underlying reorientation is a Gaussian or Markovian process, respectively. 
\revise{Other variants in this family such as active Ornstein-Uhlenbeck particles (AOUP)~\cite{fodor2016far,fodor2018statistical} have been proposed as well.
}
The directed swimming motion persists at short times ($t \ll \tau$) during which one observes ballistic motion of an individual particle. At times much larger than $\tau$, the swimming motion becomes a random walk characterized by the so-called swim diffusivity, $D^\mathrm{swim} \sim U_0^2/\tau$.

However, 
many active biological systems such as albatrosses~\cite{viswanathan1996levy}, bumblebees and deer~\cite{edwards2007revisiting}, bacteria~\cite{figueroa2020coli,figueroa20203d,huo2021swimming,korobkova2004molecular}, soil ameoba~\cite{levandowsky1997random}, fungi~\cite{asenova2016optimal}, T cells~\cite{harris2012generalized}, as well as humans~\cite{gonzalez2008understanding,brockmann2006scaling,brockmann2013hidden}, are known to exhibit super-diffusive behavior at long times described by L\'{e}vy processes~\cite{shlesinger1986levy,shlesinger1986levy2,zaburdaev2015levy}.  
\revise{Hydrodynamics of active suspensions can also induce L\'evy motion of passive tracers~\cite{kanazawa2020loopy}.
}
In particular, recent experimental works have shown that, under certain conditions, \textit{Escherichia coli (E. coli)} can travel comparatively large distances that exceed predictions from the RTP model with exponentially-distributed run-times~\cite{figueroa2020coli,huo2021swimming}. Instead, a power-law distribution of run-times that leads to  L\'{e}vy motion has been shown to explain the observed persistent motion~\cite{huo2021swimming}. L\'{e}vy motions are understood as a more advantageous foraging or search strategy than Brownian motion in many settings~\cite{viswanathan1999optimizing,lomholt2008levy,benichou2011intermittent,viswanathan2011physics,reynolds2009levy,humphries2012foraging}, leading to applications in robotics and artificial intelligence for better search pattern design~\cite{nurzaman2009yuragi,sutantyo2010multi,nurzaman2010adaptive,estrada2020interacting}. 

As pointed out by 
\citet{zaburdaev2015levy}, when discussing L\'evy motion, it is important to distinguish L\'{e}vy \textit{walks} from the earlier mathematical model of L\'{e}vy \textit{flights}. The latter, by definition, are jump processes with unbounded velocity magnitudes. At each time interval, a spatial jump is drawn from an appropriate heavy-tailed distribution, leading to discontinuous-in-time paths in the continuum limit. Such models violate Einstein's locality principle in physics, and the nature of instantaneous jumps make it impractical, if not impossible, to deal with in interacting systems.
\revise{The resolution of instantaneous jumps in overdamped Langevin dynamics (of ABP, RTP, AOUP) was discussed  by~\citet{fodor2018non}, while active matter models based on L\'evy flights were proposed in~\citet{cairoli2019active}.
}
On the other hand, L\'evy walks involve particles that change direction at random times, but travel in each direction with bounded (e.g., constant) velocity for a persistent time sampled from a heavy-tailed distribution. Such motion can be thought of as interpolation between ballistic motion and Brownian motion. 
Fractional Brownian motion (FBM) has also been studied~\cite{decreusefond1999stochastic,biagini2008stochastic,vojta2019probability,vojta2020reflected} in parallel to the L\'evy walk model. FBM shares many similar behaviors with L\'evy walks, but it is constructed from correlated noise and falls within the framework of Gaussian processes.

In general, the continuum limit of L\'{e}vy processes is described by fractional-order calculus.
Since the first attempt on fractional calculus by Leibniz in 1695, mathematicians have developed this idea into a rich field where  definitions for differentiation and integration abound~\cite{de2014review,teodoro2019review}, generalizing the concept of integer-order calculus in different ways. A fundamental connection between fractional calculus and L\'{e}vy processes is suggested by the generalization of the central limit theorem (CLT)~\cite{levy1954theorie,gnedenko1954limit,meerschaert2019stochastic,meerschaert2001limit}, in which the average of independent, identically distributed (i.i.d.) random variables with possibly infinite variance but  power-law (Paretian) tails converge to $\alpha$-stable  distributions\change{~\cite{gnedenko1954limit,nolan2020univariate}}. The Gaussian distribution is the special case of finite variance statistics, corresponding to the Wiener (Brownian) process. Comparing the characteristic function of an $\alpha$-stable distribution with a Gaussian one then hints at the more general formula of a fractional derivative in Fourier space, leading to the Riesz fractional derivative. The development of the continuous-time random walk (CTRW) framework~\cite{montroll1965random} and later L\'{e}vy walks as a more physical alternative to L\'{e}vy flights~\cite{zaburdaev2015levy} lead a wide variety of proposed fractional-order models, especially in bounded domains. However, in many studies, the form of fractional PDEs were hypothesized based on intuition or convenience rather than a rigorous derivation, and explicit treatment of boundaries are often left out.

In many biological and industrial processes, such as infection by motile bacteria or the formation of biofilms, active swimmers are confined by a boundary. An active particle with a constant swim speed that collides with a wall keeps pushing against the wall and sliding along it, and is able to swim away only after a reorientation event that turns it into the bulk. This steric interaction leads to a boundary \change{accumulation layer} of active particles at the wall, whereas passive Brownian particles exhibit a uniform density distribution in the entire domain. This behavior has been exploited to design several interesting mechanical systems including microscale gears powered by motile bacteria suspensions~\cite{Sokolov2010} and mechanical traps used to collect bacteria from the surrounding fluid~\cite{di2017deployable,wondraczek2019artificial,gutekunst20193d}. In the presence of flow, such as in the human urinary tract and medical catheters, the \change{vorticity of the flow orients the particles in the accumulation boundary layer to pointing upstream, and this} upstream swimming of bacteria often leads to infection~\cite{peng2020upstream,daddi2020tuning}. Nano-robots~\cite{soto2020medical} usually work in confined geometries, especially in porous networks such as human blood vessels. 
Better understanding of the density distribution in these bounded geometries can assist the design of devices and robots, further advancing environmental or medical engineering \change{applications}. Previous literature has considered confinement effects of ABPs~\cite{kjeldbjerg2021theory} and RTPs~\cite{angelani2017confined,ezhilan2015distribution}, where the non-uniform steady state density across a channel is due to boundary layer accumulations.

Motivated by the recent biological and microfluidic experiments~\cite{korobkova2004molecular,figueroa2020coli,figueroa20203d,huo2021swimming}, as well as progress in derivations of fractional PDE's from stochastic kinetics of L\'{e}vy walks in unbounded domains~\cite{perthame2018fractional,estrada2018fractional}, here we propose a model of micro-swimmers that perform L\'{e}vy walks in bounded domains, and refer to these swimmers as active L\'{e}vy swimmers. 
Whereas the L\'{e}vy walk model only concerns the trajectory of a single walker, our active L\'{e}vy swimmer model can incorporate physical interactions among the swimmers at finite packing density and with other objects as well as with the confining geometry. 
Moreover, while elastic reflecting boundary conditions have been considered for L\'evy walks in 1D~\cite{dybiec2017levy}, leading to uniform bulk density distributions at steady-state, we impose hard sphere interactions that are widely used to model active swimmers, leading to sliding no-flux boundary conditions.  

In this first work, we consider the ideal case where the swimmers do not interact with each other, but only with the walls. \change{
We derive a full set of transport equations for the probability density function of the L\'evy swimmers, and analyze the limits of weak and strong confinement.
In the weak confinement limit, we show that these equations reduce to a fractional-order diffusion equation with sliding no-flux boundary conditions for the number density.
We report U-shaped bulk density distributions at steady state, in contrast to the constant-in-the-bulk profiles of ABPs and normal RTPs.
Our theory provides a foundation for studies of interacting active L\'evy swimmers.}

\section{The model}
\label{sec:model}
\begin{figure}[t]
\centering
\includegraphics[width=0.8\columnwidth]{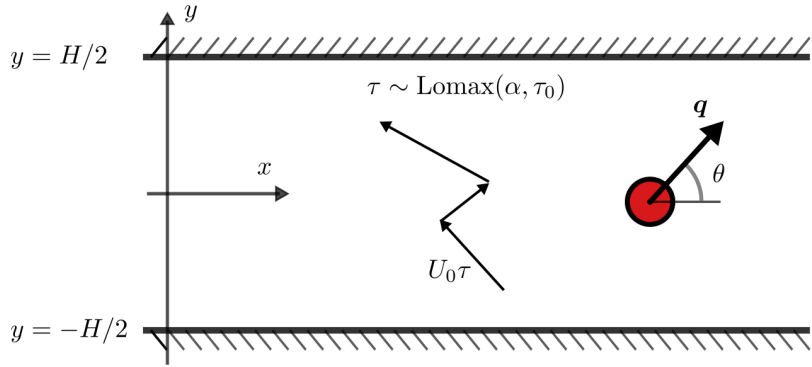}
\centering
\caption{Schematic of a L\'{e}vy swimmer confined in a 2D channel. The run-time $\tau$ follows a Lomax distribution, featuring a power-law tail. When the swimmer encounters a wall, it will maintain its orientation and slide along the wall before tumbling.
\label{fig:schematic}}
\end{figure}

As shown in Figure~\ref{fig:schematic}, an active L\'{e}vy swimmer is a sphere that propels itself with a constant swim force. We neglect hydrodynamic interactions. In Stokes flows, inertia is negligible and this results in a constant swim speed ${U}_0\bm{q}$ when not interacting with other swimmers or with a wall, where $\bm{q}$ is the unit vector representing the orientation. The swimmer changes the orientation of the swim force -- hence the velocity direction $\bm{q}$ -- stochastically, a behavior referred to as tumbling. The run-time $\tau$ between two consecutive tumbles is sampled from a type-II Pareto distribution (also known as the Lomax distribution~\cite{lomax1954business, albers2018exact})
\begin{equation}\label{eqn:lomax}
\psi(\tau)=\frac{\alpha \tau_0^\alpha}{(\tau+\tau_0)^{\alpha+1}},
\end{equation}
where $\tau_0$ is a characteristic timescale.
It is known that when $0<\alpha<1$ the mean run-time $\tau_m$ diverges and L\'{e}vy walks in this regime display weak ergodicity breaking~\cite{albers2018exact}, so that the time average differs from the ensemble average. 
We will restrict our model of L\'{e}vy swimmers to $1<\alpha<2$; in this case, the mean run-time $\tau_m$ is finite and related to the characteristic timescale via 
\begin{equation}
\tau_m=\frac{\tau_0}{\alpha-1}.
\end{equation}
 The $\alpha>2$ regime is expected to converge to the Brownian limit since the variance $\sigma^2$ is finite and is given by $\sigma^2 = {\tau_m^2 \alpha}/{(\alpha-2)}$.
We compare the results from L\'{e}vy swimmers with those of $\alpha=3>2$, using the later case to represent Gaussian models by appealing to the CLT. 
When a L\'evy swimmer collides with another swimmer, it is constrained by hard sphere repulsions. When it collides with a wall, it is constrained by hard wall repulsion and remains free to slide along the wall, in a similar way to a Skorokhod boundary condition for Brownian motion~\cite{skorokhod1961stochastic,dupuis1999convex}.
Specific details of the particle simulations used in this article are given in \ref{sec:simulation_details}

For simplicity, we perform analysis for non-interacting L\'{e}vy swimmers confined by walls of a channel. We analyze the density distribution and forces exerted on the walls through kinetic equations, and rigorously prove a fractional diffusion equation governing the steady state density in the weak confinement limit by extending the works of Refs.~\cite{frank2016fractional,perthame2018fractional,estrada2018fractional}. Our results are compared to particle simulations. In Section \ref{sec:weak-limit}, we show that at steady state, L\'{e}vy swimmers maintain a U-shaped {\it bulk} distribution even at weak confinement, in contrast to the accumulation {\it boundary layer} effect of ABPs and RTPs, which scales with activity, manifesting the fundamental difference between L\'{e}vy processes (fractional-order) and Gaussian processes (integer-order).

\section{Transport equations for ideal L\'{e}vy swimmers in a channel}
The stochastic process of L\'{e}vy swimmers can be described by a set of transport equations for the particle probability distribution. There are two features that are \revise{noteworthy compared with} the ABP\revise{/RTP} model. First, there is no explicit translational diffusion; as a result, we expect singular accumulation of the particles on the boundaries, \change{as would occur even} for normal RTP with exponentially distributed run-times~\cite{ezhilan2015distribution}. \revise{Second, 
both ABP and normal RTP are essentially Markovian and lose memory at the long-time limit.} However, this Markovian property is lost for L\'{e}vy swimmers given the slow power-law decay of their run-time distribution. Hence, we have to explicitly keep track of the current run-time variable of the L\'evy swimmer, denoted by $\tau$~\cite{alt1980biased,estrada2018fractional}. With these in mind, we can write down the kinetics in 2D. The distribution of particles inside the channel of width $H$ is governed by
\begin{equation}
\left(\frac{\partial}{\partial\tau} + \frac{\partial}{\partial t} + U_0\bm{q}\cdot\nabla \right) P(x,y,t,\tau,\theta) = -\beta(\tau) P(x,y,t,\tau,\theta), \label{eqn:micro-bulk} \\
\end{equation}
where $\bm{q}=(\cos\theta,\sin\theta)$ is the unit vector for particle orientation, and $-H/2<y<H/2$. The term $\beta(\tau)$ is the tumbling rate function, which is generally a function of the current run-time $\tau$. 
For active L\'{e}vy swimmers it is given by
\begin{equation}
\beta(\tau)=\frac{\alpha+1}{\tau+\tau_0},
\end{equation}
\change{
In the case of normal RTP, $\beta=1/\tau_m$, a constant, and equation~\eqref{eqn:micro-bulk} reduces to the Smoluchowski equation~\eqref{eqn:normal_rtp_equation}, as shown below. }

This equation can be understood by considering an infinitesimal time interval $dt$: the probability density $P$ is shifted in space-time by the material derivative $(\partial/\partial t + U_0\bm{q}\cdot\nabla) P dt$. For all swimmers that did not tumble during $dt$, their current run-time increases by $d\tau=dt$, hence the term $\partial P/\partial \tau $ in equation \eqref{eqn:micro-bulk}. A fraction of the swimmers $\beta(\tau) P dt$ tumbled to other directions during $dt$, giving the sink term on the RHS of equation \eqref{eqn:micro-bulk}. All of these swimmers that just tumbled now have their current run-time renewed to be $\tau=0$, hence giving the initial condition for $\tau$,
\begin{equation}
P(x,y,t,0,\theta) = \frac{1}{2\pi} \int_0^t \beta(\tau') d\tau' \int_0^{2\pi}  P(x,y,t,\tau',\theta')d\theta'
\end{equation}
Similarly, for the probability density on the upper wall, we have
\begin{equation}
\frac{\partial \phi^+}{\partial \tau} + \frac{\partial \phi^+}{\partial t}+ U_0 \cos\theta^+ \frac{\partial \phi^+}{\partial x}  = -\beta(\tau) \phi^+ + j_b^+ \label{eqn:micro-wall},
\end{equation}
where $\phi^+$ is short for the number density $\phi$ on the wall at the upper wall $y = H/2$,
\begin{equation}
\phi^+ = \phi(x,y=H/2,t,\tau,\theta).
\end{equation}
The additional source term $j_b^+$ is the net flux of swimmers going from the bulk into the upper wall at $y=H/2$
\begin{equation}
j_b^+ = U_0 P(x,y=H/2,t,\tau,\theta^+) \sin(\theta^+) \label{eqn:micro-flux-into-wall}.
\end{equation}
The equation for the lower wall at $y=-H/2$ is symmetric to that of the upper wall and is omitted here. The superscripts on $\theta^\pm$ is a reminder of $0<\theta^+<\pi$ and $-\pi<\theta^-<0$.

The bulk density flux close to the wall is coupled with the swimmer flux tumbling from the wall through
\begin{equation}
\int_0^t \beta(\tau)d\tau \frac{1}{2\pi} \int_0^\pi d\theta' \phi^+  = \phi^+(x,t,0,\theta^+)
= -P(x,y=H/2,t,0,\theta^-)U_0\sin\theta^- \label{eqn:micro-flux-out-wall}.
\end{equation}
In this setting, the only parameter for RTPs is the strength of confinement $\ell/H$, where $\ell = U_0\tau_m$ is the average run length.
For L\'{e}vy swimmers there is another parameter, the power-law index $\alpha$ of run-time distributions, independent of the average run-time $\tau_m$.
At steady state, equations \eqref{eqn:micro-bulk}--\eqref{eqn:micro-flux-out-wall} can be further simplified; see  \ref{sec:appendix_a}.

We note that one can recover the equations for normal RTPs from the above general kinetic equations (3-6). In that case $\beta(\tau)=\beta=1/\tau_0$ is a constant, hence integration over $\tau$ is trivial and gives
\begin{align}\label{eqn:normal_rtp_equation}
\begin{split}
\left(\frac{\partial }{\partial t} + U_0\bm{q}\cdot\nabla \right) p(\bm{r},t,\theta) & = 
\frac{1}{2\pi}\int_0^{2\pi}  p(\bm{r},t,\theta')d\theta' -\beta p(\bm{r},t,\theta)
- P(\bm{r}-U_0t\bm{q},0,0,\theta) e^{-\beta t} \\
& = \mathcal{L}p - P(\bm{r}-U_0t\bm{q},0,0,\theta) e^{-\beta t},
\end{split}
\end{align}
where 
\begin{equation}
p(\bm{r},t,\theta)=\int_0^t P(\bm{r},t,\theta,\tau)d\tau,
\end{equation}
is the reduced probability density,
and the turning operator $\mathcal{L}$ is defined~\cite{othmer2000diffusion,plaza2019derivation}
by
\begin{equation}
\mathcal{L}p=\frac{1}{2\pi}\int_0^{2\pi}  p(\bm{r},t,\theta')d\theta' -\beta p(\bm{r},t,\theta).
\end{equation}
In the long time limit, the last term of equation \eqref{eqn:normal_rtp_equation} represents the exponentially vanishing initial condition, and is usually omitted in multiscale analysis of RTPs.

\section{Limit of strong confinement: no time to tumble in the bulk\label{sec:strong-limit}}

When $\ell/H\gg 1$, the confinement is very strong, so the swimmers barely tumble in the bulk of the channel. Therefore, one expects the swimmers to contribute equally to the bulk density profile at each point in space since they merely cross the channel width with a constant speed.

\begin{figure}[t]
\centering
\includegraphics[width=\columnwidth]{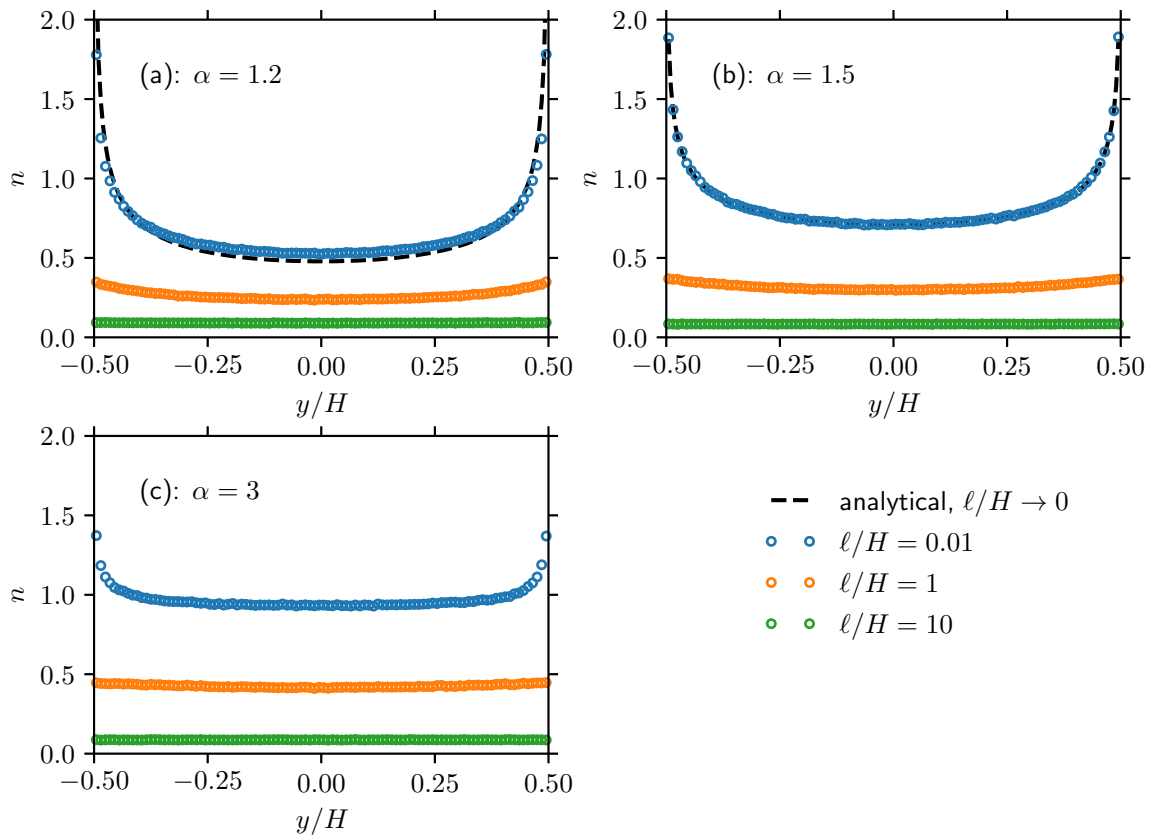}
\centering
\caption{Simulated bulk density profile $n$ of swimmers with (\emph{a}) $\alpha=1.2$, (\emph{b}) $\alpha=1.5$ and (\emph{c}) $\alpha=3.0$ in a 1D channel. The strong confinement limit of (\emph{a}--\emph{c}) and the weak confinement limit of (\emph{a}--\emph{b}) agree with the analysis in Sections~\ref{sec:strong-limit} and \ref{sec:weak-limit}. 
In contrast to (\emph{a}--\emph{b}), (\emph{c}) reveals convergence to the Gaussian limit with a constant bulk density and a thin boundary layer for $\ell/H\ll 1$, due to the finite variance of Pareto distribution with $\alpha=3>2$. This is consistent with the RTPs results reported in Ref.~\cite{ezhilan2015distribution}.
In all subfigures (\emph{a}--\emph{c}), when normalizing the densities, only swimmers in the bulk, but not on the wall, are included.
The swimmers accumulated on the walls are not shown here but they give rise to the forces on the wall shown in Figure \ref{fig:force-polarorder} (a).
\label{fig:1D-density}}
\end{figure}

\begin{figure}[t]
\includegraphics[width=\columnwidth]{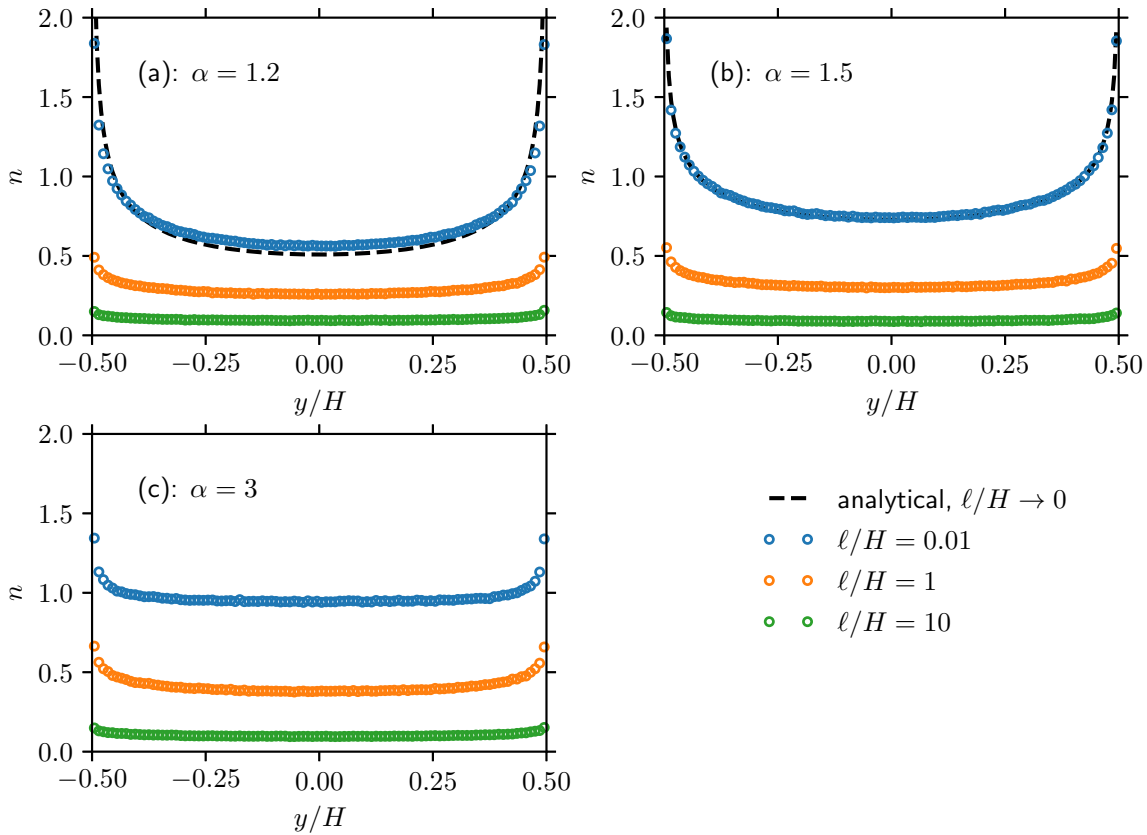}
\caption{Simulated bulk density profile $n$ of swimmers with (\emph{a}) $\alpha=1.2$ (\emph{b}) $\alpha=1.5$ and (\emph{c}) $\alpha=3.0$ across a 2D channel. The weak confinement limit of (\emph{c}) converges to the CLT prediction of constant in the bulk, proving that the U-shape density profile in (\emph{c}) for intermediate $\ell/H$ is only a boundary layer effect~\cite{ezhilan2015distribution}, while in (\emph{a}--\emph{b}) it is a bulk property of the superdiffusive behavior.
When normalizing the densities, only swimmers in the bulk but not on the wall are included.
The swimmers accumulated on the walls are not shown here but they give rise to the forces on the wall shown in Figure \ref{fig:force-polarorder} (b).
\label{fig:2D-density}
}
\end{figure}

In 1D this is asymptotically true for both L\'{e}vy and normal swimmers when their average run-time $\tau_m$ is longer than the channel crossing time $t_c=H/{U_0}$. 
As shown in Figure~\ref{fig:1D-density}, the simulated bulk density becomes almost flat \revise{in the strong confinement limit for which} $\ell/H>1$. 
\revise{In Figure~\ref{fig:1D-density} (as in Figure~\ref{fig:2D-density}), each simulation curve integrated over the interior of the channel gives the corresponding bulk fraction of swimmers, not including the singular accumulation on the walls.}
In \revise{the strong confinement} limit the fraction of swimmers accumulated exactly on the wall can be estimated by time average of the individual trajectories, noting that we restrict the power-law index $\alpha>1$ so that time average equals the ensemble average~\cite{albers2018exact}. In 1D, the time spent inside the bulk is a constant $t_c$. The fraction of swimmers in the bulk is then
$f_\mathrm{bulk}={H}/\ell$. Since in 1D the swimmer only takes orientations of either to the left or to the right, this partitioning between bulk and boundary immediately translates into the force on the wall, as in equation~\eqref{eqn:force-wall}.
\revise{ As the confinement strength $\ell/H$ increases, larger fractions of swimmers are stuck on the wall. Hence, the total bulk fraction shown on the plots decreases from blue ($\ell/H=0.01$) to orange ($\ell/H=1$) to green ($\ell/H=10$) curves.}

\begin{figure}[t]
\includegraphics[width=\columnwidth]{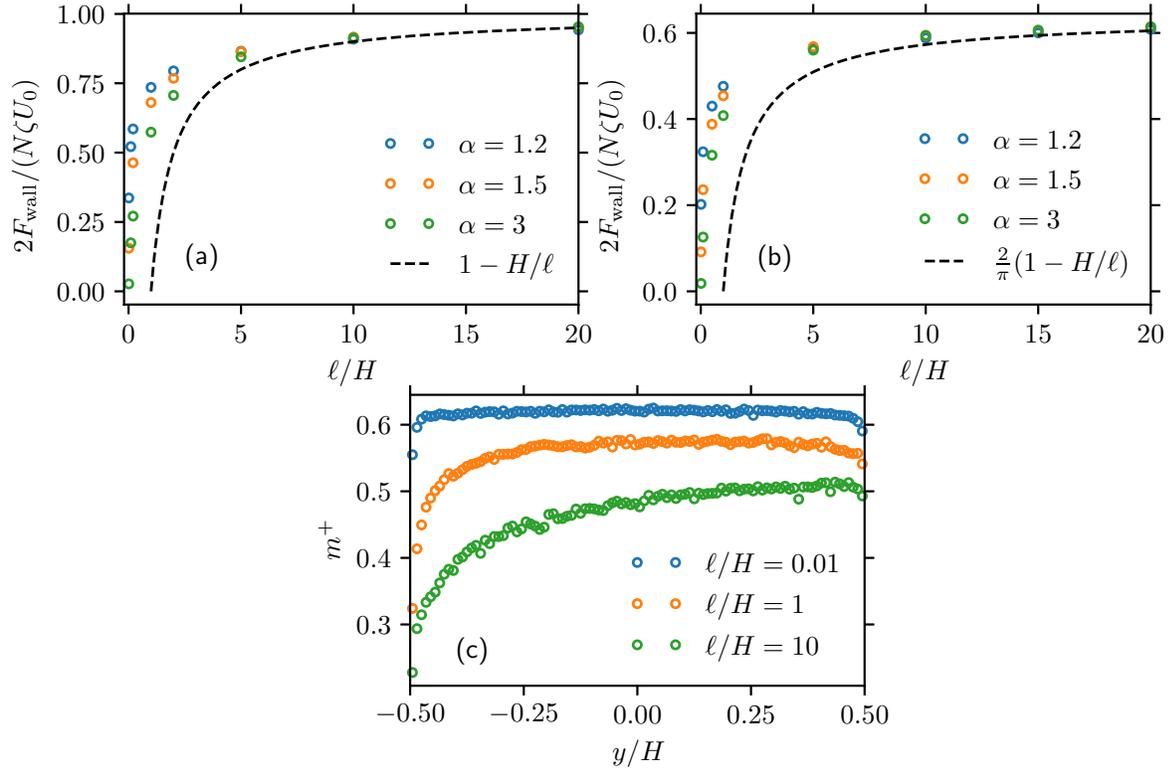}
\caption{Forces of swimmers with $\alpha\in\{1.2, 1.5, 3\}$ exerted on the wall in (a) 1D and (b) 2D channels. The asymptotes are from the strong confinement limit analysis in Section \ref{sec:strong-limit}, which is expected to be more accurate as $\ell/H$ becomes larger. (c) Distribution of the polarization $m^+$ [see equation \eqref{eqn:polarization}] of 2D L\'{e}vy swimmers  with $\alpha=1.2$ pointing towards the upper wall.
\label{fig:force-polarorder}}
\end{figure}

In 2D, the analysis remains true in the same asymptotic limit, with the slight modification that the swimmer orientation now ranges in $[0,2\pi)$. At strong confinement, this leads to an $O((H/\ell)^2)$ correction  to the average channel crossing time.
The polar order distribution on the wall is modified by a prefactor of \revise{$\frac{1}{\pi}\int_0^{\pi}\sin(\theta)d\theta = 2/\pi$}. The force exerted on a wall of the channel is therefore 
\begin{equation}
F_\mathrm{wall} = \frac{C_d}{2} \left( 1 - \frac{H}{\ell} \right) N\zeta U_0
\label{eqn:force-wall}
\end{equation}
where $N$ is the number of swimmers confined inside the channel,
$\zeta U_0$ is their swim force, and $C_d=1$ for 1D  and $C_d=2/\pi$ for 2D.
As show in \Cref{fig:1D-density,fig:2D-density}, the profiles for $\ell/H=10$ have almost converged to uniform density distributions, confirming the above analysis.
Each simulation curve integrated over the domain gives the corresponding bulk number of swimmers, not including the singular accumulation on the walls.
As shown in Figure~\ref{fig:force-polarorder} (a), the force asymptotes represented by equation \eqref{eqn:force-wall} agree well with the results of our particle simulations for large $\ell/H$.
\revise{The same figure illustrates how the singular accumulation at the walls, quantified by $F_\mathrm{wall}$, varies with $\ell/H$. 
}

\section{Limit of weak confinement: fractional diffusion\label{sec:weak-limit}}
When $\ell/H\ll 1$, we expect that the swimmers in the bulk do not `see' the boundary of the domain easily, and so can be modeled by the same equations as swimmers in free space, but with boundary conditions. In this section, we show this gives rise to a fractional diffusion equation involving the Riesz definition of fractional derivatives in finite domains together with a no-flux boundary condition recently introduced by \citet{kelly2019boundary} and \citet{baeumer2018reprint}.

In this limit, at times $t$ much larger than $\tau_m$, the system reaches a superdiffusive regime. 
This permits a multiscale analysis using a Hilbert expansion technique~\cite{ellis1973chapman} to coarse-grain over the short time $\tau$ of ballistic motion. To prepare for this analysis, 
the kinetic equation~\eqref{eqn:micro-bulk} 
can be cast in Fourier space as
\begin{equation}
\left(\frac{\partial}{\partial\tau} + \frac{\partial}{\partial t} + U_0\bm{q}\cdot (i\bm{\xi}) \right) \hat{P}(\bm{\xi},t,\tau,\bm{q}) = -\beta(\tau) \hat{P}(\bm{\xi},t,\tau,\bm{q}); \label{eqn:micro-bulk-fourier} \\
\end{equation}
here, we define the Fourier and inverse Fourier transforms, respectively, as
\begin{align}
\mathcal{F}[u](\bm{\xi}) 
&= \frac{1}{(2\pi)^{d/2}} \int e^{-i \bm{r} \cdot \bm{\xi}} u(\bm{r}) d\bm{r}, \\
\mathcal{F}^{-1}[\hat{u}](\bm{r})
&= \frac{1}{(2\pi)^{d/2}} \int e^{i \bm{r} \cdot \bm{\xi}} \hat{u}(\bm{\xi}) d\bm{\xi},
\end{align}
with the shorthand $\hat{u} = \mathcal{F}[u]$.
We then integrate equation~\eqref{eqn:micro-bulk-fourier} over $\tau$ for the reduced probability density
\begin{equation}
\hat{p}(\bm{\xi},t,\bm{q})=\int_0^{t}\hat{P}(\bm{\xi},t,\tau,\bm{q})d\tau.
\end{equation}
In Fourier space, this leads (see equation (14) of Ref.~\cite{estrada2018fractional}) to
\begin{equation}\label{eqn:phat}
\left(\frac{\partial}{\partial t} + U_0\bm{q}\cdot (i\bm{\xi}) \right) \hat{p}(\bm{\xi},t,\bm{q}) = (\mathcal{T} - \mathcal{I}) \int_0^t \beta(\tau)\hat{P} d\tau 
\end{equation}
where $\mathcal{I}$ is the identity operator and $\mathcal{T}[f(\bm{q})]=\int_S f(\bm{q}')d\bm{q}'$ is the turn angle operator, with $S$ denoting the unit sphere in $\mathbb{R}^d$. 

Our goal is to derive the governing equation for the number density of swimmers
\begin{equation}
n(\bm{r},t):= \int_S  p(\bm{r},t,\bm{q}) d\bm{q}.
\end{equation}
The conservation of number density is expressed by 
\begin{equation}\label{eqn:conservation}
\frac{\partial n}{\partial t} + \nabla \cdot {\bm{j}} = 0,
\end{equation}
where $\bm{j}$ is the flux, which is defined as
\begin{equation}
\bm{j}(\bm{r},t) := \int_S U_0\bm{q} p(\bm{r},t,\bm{q}) d\bm{q}.
\end{equation}

\subsection{Derivation of fractional-order flux via Hilbert Expansion}\label{sec:derivation_free_space}
Refs.~\cite{frank2016fractional,estrada2018fractional,perthame2018fractional} show that by assuming a superdiffusive power-law scaling for the leading order equation obtained in the Hilbert expansion analysis of equation \eqref{eqn:phat}, a fractional-order governing equation can be derived for the number density $n$ in free space. We summarize their derivation below, with the modification that the equations are represented in Fourier space. 
By expressing the derivations of \citet{estrada2018fractional} in Fourier space, their formal expansions in powers of differential operators become expansions in powers of scalars, which are well-defined. In turn, this justifies the definition of the specific fractional-order gradient operator in their final results. This operator was introduced by \citet{meerschaert2006fractional} as
\begin{equation}\label{eqn:fractional_gradient_fourier}
\mathcal{F} \big[\nabla^{\alpha-1} u\big](\bm{\xi}) = 
\left[ \int_{S} \bm{q} (i\bm{\xi} \cdot \bm{q})^{\alpha-1} d\bm{q} \right] \hat{u}(\bm{\xi});
\end{equation}
Clarification of the specific fractional-order gradient operator is critical, as there are several notions of gradient in fractional calculus that are not equivalent \cite{d2020unified,vsilhavy2020fractional}.

Starting from equation \eqref{eqn:phat},
\citet{estrada2018fractional} showed
$p(\bm{r},t,\bm{q})$ can be formally expanded in terms of $\bm{q}\cdot\nabla$, which is rigorous in Fourier space since $\bm{q}\cdot\nabla$ transforms into the scalar $\bm{q}\cdot(i\bm{\xi})$. The leading order of their expansion led to equation (46) in their article, which in Fourier space, and in the absense of external source terms, is represented as
\begin{equation}
\int_S \bm{q} \frac{(\alpha-1)}{\tau_0}\bm{q}\cdot \hat{\bm{j}} d\bm{q} = \int_S \bm{q} \tau_0^{\alpha-2} (1-\alpha)^2 \Gamma(1-\alpha) U_0^{\alpha} (\bm{q}\cdot(i\bm{\xi}))^{\alpha-1} \hat{n} d\bm{q}.
\end{equation}
In our case, this expression is simplified due to the uniform distribution of orientation angle after tumbling events.
\citet{estrada2018fractional} then show that
\begin{equation}
\int_S \bm{q} \frac{(\alpha-1)}{\tau_0}\bm{q}\cdot \hat{\bm{j}} d\bm{q} 
= \frac{(\alpha-1)}{\tau_0} \hat{\bm{j}}.
\end{equation}
As a result of the last two equations, the Fourier transform of the flux $\hat{\bm{j}}$ can be related to the Fourier transform of the number density $\hat{n}$,
\begin{equation}
\hat{\bm{j}}=
\frac{\pi (\alpha-1)\tau_0^{\alpha-1}U_0^{\alpha}}{\sin(\pi\alpha)\Gamma(\alpha)}
\int_S \bm{q}  (\bm{q}\cdot i\bm{\xi})^{\alpha-1} \hat{n} d\bm{q}.
\end{equation}
This coincides with the definition of the fractional gradient \eqref{eqn:fractional_gradient_fourier}, and 
when transformed to real space gives
\begin{equation}
{\bm{j}} = 
\frac{\pi (\alpha-1)\tau_0^{\alpha-1}U_0^{\alpha}}{\sin(\pi\alpha)\Gamma(\alpha)}
\nabla^{\alpha-1} n.
\end{equation}

Inserting the flux expression into the conservation equation~\eqref{eqn:conservation}, and using the property \cite{meerschaert2006fractional}
\begin{equation}
\nabla \cdot \nabla^{\alpha - 1} = -(-\Delta)^{\alpha/2}, 
\end{equation}
this finally leads to
\begin{equation}
\frac{\partial{n}}{\partial t}
-C_\alpha (-\Delta)^{\alpha/2} n = 0,
\label{eqn:fractional-realspace}
\end{equation}
where $C_\alpha=
{\pi (\alpha-1)\tau_0^{\alpha-1}U_0^{\alpha}}/\left[{\sin(\pi\alpha)\Gamma(\alpha)}\right]$ and the fractional Laplacian is defined as
\begin{equation}
\label{eqn:fractional-laplacian-def}
-(\Delta)^{\alpha/2}n(\bm{r})
=
\frac{2^\alpha \Gamma(\frac{\alpha}{2} + \frac{d}{2})}{\pi^{d/2}|\Gamma(-\alpha/2)|}
\int_{\mathbb{R}^d} \frac{n(\bm{r}) - n(\bm{r}')}{|\bm{r}-\bm{r}'|^{d+\alpha}} d\bm{r}'.
\end{equation}

It is possible to write
\begin{equation}
-(-\Delta)^{\alpha/2}n = \Delta I_{2-\alpha} n
\end{equation}
where $I_{2-\alpha}$ is a Riesz potential of order $2-\alpha$ applied to $n$, the inverse Fourier transform of which is 
 $\vert \bm{\xi} \vert^{-(2-\alpha)}  \hat{n}$. As a result, one can write equation~\eqref{eqn:fractional-realspace} as
\begin{equation}
\begin{split}
\frac{\partial{n}(\bm{r},t)}{\partial t} 
&=
- C_\alpha 
\ \Delta
\int_{\mathbb{R}^d} \frac{n(\bm{r}',t)}
{\vert \bm{r}-\bm{r}'\vert^{1-(2-\alpha)}} d\bm{r}'.
\end{split}
\end{equation}
It is possible then to restrict integration of $\bm{r}'$ to the channel. 
We take this step in the following section after considering the steady state of equation \eqref{eqn:fractional-realspace}. 

We remark here on equation~\eqref{eqn:fractional-realspace}: 
It is known that this fractional diffusion governs the probability distribution of a L\'evy \textit{flight} process in free space, with the solution exhibiting unbounded support due to the unbounded jumps inherent to the flight paths. Nevertheless, it has been derived as a first-order governing equation for the L\'evy walk model that describes non-interacting active L\'evy swimmers.
\citet{zaburdaev2015levy} has demonstrated that the bulk density profile of L\'evy walkers in free space is approximated by a L\'evy stable distribution supported between two ballistic peaks that propagate with constant velocity. This  result is consistent with the derivation of equation \eqref{eqn:fractional-realspace} as a leading-order equation for the distribution of L\'evy walkers. Below, we carefully show how boundary conditions can be utilized to constrain the solution of equation \eqref{eqn:fractional-realspace}
to describe the properties of the active L\'evy swimmers introduced in Section \ref{sec:model}.  
We also note that a Hilbert expansion analysis of normal run-and-tumble particles in 1D results in a telegraph equation~\cite{mckean1967chapman,ellis1973chapman}, which yields a propagating front with finite speed due to a second order time derivative term. It is possible that a more careful multiscale analysis for L\'evy walks, which includes higher order effects, may yield an additional hyperbolic term in equation \eqref{eqn:fractional-realspace} to properly characterize the ballistic front.

\subsection{Zero exterior condition and no-flux boundary condition for the channel geometry}
In the previous section, we reviewed that a fractional diffusion equation governs the number density $n$. Here, we clarify the boundary conditions required to close the description of the active L\'evy swimmers confined in the channel. 
Due to their discontinuous-in-time paths, 
boundary conditions for L\'{e}vy processes raise complicated issues, often requiring exterior conditions for their governing equations instead \change{of} standard boundary conditions~\cite{lischke2018fractional}.

According to the hard-wall interactions inherent in our model of active L\'evy swimmers described in Section \ref{sec:model}, particles cannot pass through boundaries of the channel, which manifests in two properties. First, the density vanishes, 
\begin{equation}
P(\bm{r},t,\tau,\bm{q})=0,
\end{equation}
for $\bm{r}$ outside of the channel, which implies that 
\begin{equation}\label{eqn:exterior_condition}
n(\bm{r},t)=0.
\end{equation}
Second, the flux vanishes, 
\begin{equation}\label{eqn:flux_condition}
\bm{j}(\bm{r},t) = 0,
\end{equation}
when $\bm{r}$ lies on the channel wall. For a classical diffusion equation, only the second condition is required, as it provides a Neumann boundary condition which determines the solution up to a constant. However, the fractional Laplacian \eqref{eqn:fractional-laplacian-def}, due to its nonlocal nature, requires an exterior condition represented by equation \eqref{eqn:exterior_condition} to be well-defined~\change{\cite{lischke2018fractional}}. Below, we will use conditions \eqref{eqn:exterior_condition} and \eqref{eqn:flux_condition} in distinct ways to derive the steady state solution to equation \eqref{eqn:fractional-realspace}. 

\subsection{Steady state in 1D}\label{sec:1d_steady_state}
In 1D, the fractional Laplacian in equation \eqref{eqn:fractional-realspace} reduces to the the Riesz (or the Riesz-Feller) fractional derivative in one dimension \cite{kelly2019boundary}. This operator can be written for $1 < \alpha < 2$ as
\begin{equation}\label{eqn:riesz_deriative}
\frac{\partial^\alpha n(y)}{\partial\vert y \vert^\alpha}
=
\frac{C_\alpha}{\Gamma(2-\alpha)}
\frac{d^2}{dy^2}
\int_{-\infty}^{\infty}
\frac{n(x)}{|y-x|^{\alpha-1}} dx.
\end{equation}
Using the exterior condition \eqref{eqn:flux_condition}, which in 1D implies that $n(y) = 0$ for $y \not\in [-H/2, H/2]$, this reduces to
\begin{equation}\label{eqn:finite-domain-riesz-operator}
\frac{\partial^\alpha n(y)}{\partial\vert y \vert^\alpha}
=
\frac{C_\alpha}{\Gamma(2-\alpha)}
\frac{d^2}{dy^2}
\int_{-H/2}^{H/2}
\frac{n(x)}{|y-x|^{\alpha-1}} dx.
\end{equation}
Therefore, the steady state of equation \eqref{eqn:fractional-realspace} can be written
as 
\begin{equation}\label{eqn:steady_state}
0=
\frac{\partial^\alpha n(y)}{\partial\vert y \vert^\alpha}
=
\frac{C_\alpha}{\Gamma(2-\alpha)}
\frac{d^2}{dy^2}
\int_{-H/2}^{H/2}
\frac{n(x)}{|y-x|^{\alpha-1}} dx.
\end{equation}

No-flux boundary conditions for the steady state equation \eqref{eqn:steady_state}
were derived by \citet{kelly2019boundary}. They introduced the Riemann-Liouville flux, which in the symmetric case is given by 
\begin{equation}
j_{RL} =
\frac{C_\alpha}{\Gamma(2-\alpha)}
\frac{d}{dy}
\int_{-H/2}^{H/2}
\frac{n(x)}{|y-x|^{\alpha-1}} dx.
\end{equation}
This allowed them to write the steady state equation \eqref{eqn:steady_state} in conservation form, 
\begin{equation}
0 = 
\frac{dj_{RL}}{dy}
=
\frac{d}{dy}
\left(
\frac{C_\alpha}{\Gamma(2-\alpha)}
\frac{d}{dy}
\int_{-H/2}^{H/2}
\frac{n(x)}{|y-x|^{\alpha-1}} dx
\right).
\end{equation}
Comparing this equation to equation \eqref{eqn:conservation} shows that $j_{RL}$ can be identified with $\bm{j}$ in 1D, so that the no-flux boundary condition \eqref{eqn:flux_condition} implies $j_{RL} = 0$ for $y = \pm H/2$. 
\citet{kelly2019boundary} derived a general solution to equation \eqref{eqn:steady_state}, which under the no-flux condition reduces to
\begin{equation}\label{eq:steady_state_solution}
n(y) = C_0 \left(\frac{1}{4}-\frac{y^2}{H^2}\right)^{\alpha/2-1},
\end{equation}
where $C_0$ is a normalization constant determined by conservation of total number of swimmers.

As shown in \Cref{fig:1D-density}(a--b), the analytical solution agrees well with our simulation data in the weak confinement limit ($\ell/H \ll 1$). In plotting \Cref{fig:1D-density,fig:2D-density}(a--b), the analytic curve is normalized to have the same number of total swimmers as that of the bulk from the simulations. 
The fraction of swimmers accumulated on the wall push against the wall with a constant force, giving rise to the force measured in Figure \ref{fig:force-polarorder} (a).

We remark that, in 1D, equation \eqref{eqn:fractional-realspace} with the boundary conditions \eqref{eqn:exterior_condition} and \eqref{eqn:flux_condition} also governs the evolution of the density of $\alpha$-stable L\'evy flights with the `stopping' boundary condition studied by \citet{dybiec2017levy}. This condition is defined by the property that a flier which attempts to leave the interval is stopped near the corresponding endpoint until another jump is drawn from the $\alpha$-stable distribution that moves it back into the bulk. In this context, \citet{denisov2008steady} derived a steady-state solution that agrees with equation \eqref{eqn:steady_state}. However, the prescription of `stopping' boundary conditions for L\'evy flights in higher dimensions is more subtle and is not expected to agree with the sliding no-flux boundary conditions prescribed in our active L\'evy swimmer model. 
We also remark that the shape of our steady state solution is similar to that observed for reflected fractional Brownian motion~\cite{vojta2020reflected}.

\subsection{Steady state in 2D}
In 2D, we expect the steady state of the number density to be independent of the longitudinal coordinate $x$, i.e. $n(x,y)=n(y)$; see Figure \ref{fig:schematic}. In this case, we show that the fractional Laplacian of $n(x,y)$ in equation \eqref{eqn:fractional-realspace} reduces to the one-dimensional Riesz derivative in $y$. In the classical case of integer order derivatives, this follows immediately from the expansion of the Laplacian in partial derivatives; in the fractional case, it requires a more involved proof. 

In Fourier space, the fractional Laplacian can be represented as
\begin{equation}\label{eqn:fourier_laplacian}
\begin{split}
\mathcal{F}_{x,y}[-(-\Delta)^{\alpha/2} n(x,y)](\xi,\eta) 
&= \vert \xi^2 + \eta^2 \vert^{\alpha/2} \mathcal{F}_{x,y}[n](\xi,\eta) \\
&= 
\vert \xi^2 + \eta^2 \vert^{\alpha/2} \mathcal{F}_x\left[ \mathcal{F}_y[n](\eta)\right](\xi,\eta) \\
&= 
\vert \xi^2 + \eta^2 \vert^{\alpha/2}
\mathcal{F}_y[n](\eta) \delta(\xi).
\end{split}
\end{equation}
In the above $\mathcal{F}_{x}$, $\mathcal{F}_{y}$ and $\mathcal{F}_{x,y}$ denote the Fourier transform in $x$, $y$, and $(x,y)$, respectively; for an absolutely integrable function $u(x,y)$, these satisfy
\begin{equation}
\mathcal{F}_{x,y} [u] (\xi,\eta) = \mathcal{F}_{x} [ \mathcal{F}_{y} [u] (x,\eta) ] (\xi,\eta). 
\end{equation}
In equation \eqref{eqn:fourier_laplacian}, the frequency variable $\xi$ corresponds to transformation in $x$, while $\eta$ corresponds to transformation in $y$. 
The last equality in \eref{eqn:fourier_laplacian} uses the fact that $\mathcal{F}_y[n](x,\eta)=\mathcal{F}_y[n](\eta)$ does not depend on $x$, yielding a $\delta$-function Fourier transform in $x$. 
Taking the inverse Fourier transform, we obtain
\begin{equation}\label{eqn:with_1d_riesz}
\begin{split}
-(-\Delta)^{\alpha/2} n(x,y) = & \mathcal{F}^{-1}_{\xi,\eta} \left\{ \vert \xi^2 + \eta^2 \vert^{\alpha/2} \mathcal{F}_y[n] \delta(\eta) \right\} \\
= & \mathcal{F}^{-1}_\eta \left\{ \vert\eta^2\vert^{\alpha/2} \mathcal{F}_y[n](\eta)
\right\} \\
= & \frac{\partial^\alpha n(y)}{\partial\vert y \vert^\alpha}.
\end{split}
\end{equation}
The Fourier representation of the Riesz derivative used to obtain the final line can be found in, e.g., Ref. \cite{meerschaert2019stochastic}. Therefore, the steady-state equation for $n(y)$ in 2D is identical to equation \eqref{eqn:steady_state} from the 1D case, with the same exterior and no-flux boundary conditions. It has the same solution \eqref{eq:steady_state_solution}. 
The solutions are compared with particle simulations in Figure \ref{fig:2D-density}.
As in 1D, the density profiles shown only include swimmers in the bulk but not on the walls.
A swimmer accumulated on the wall pushes against the upper/lower wall, and the normal component of the force $F\sin{\theta^{\pm}}$ is balanced by the wall. Averaged over swimmers for a finite time period this gives the force measured on the wall in Figure \ref{fig:force-polarorder}(b). Although we observe no local polar order in the bulk, the polar order of only those swimmers oriented towards the upper wall is not zero and not symmetric with respect to the center line of the channel
\begin{equation}
\label{eqn:polarization}
m^+(y)=\int_0^{\pi} \cos(\theta^+) p(y,t=\infty,\theta^+) d\theta^+.
\end{equation}
This quantity 
is plotted in Figure \ref{fig:force-polarorder}(c).

\section{Comparisons between ABPs, normal RTPs and active L\'{e}vy swimmers}

Here we comment on the differences and similarities among the three species of active matter models.
First, we compare ABPs and normal RTPs. It has been shown by a mean-field treatment of phenomenological models~\cite{cates2013active,solon2015active,cates2015motility} that ABPs and RTPs behave similarly macroscopically, for example in motility-induced phase separation (MIPS). 
To further compare RTPs and ABPs, we prove that the stochastic dynamics of non-interacting ABPs leads to an exponential decay of their velocity autocorrelation function,  coinciding with that of normal RTPs. 

Without loss of generality we consider the 2D case.
The stochastic dynamics of an ABP moving in $xy$ plane can be described by the Langevin equation 
\begin{equation}\label{eqn:langevin}
\dot{\bm{v}} (t) = \eta (t) \bm{e}_z \times \bm{v}(t), 
\end{equation}
with
\begin{equation}\label{eqn:eta_definition}
\begin{split}
\left<\eta(t_1)\eta(t_2)\cdot\cdot\cdot\eta(t_{2m-1})\right> &=  0 \\
\left<\eta(t_1)\eta(t_2)\cdot\cdot\cdot\eta(t_{2m})\right> &=  \sum_{<i,j>}\sigma^{2m} \Pi_{<i,j>}\delta(t_i,t_j) \triangleq M_{2m},
\end{split}
\end{equation}
where $\bm{v}$ is the velocity vector, $\eta(t)$ a Gaussian noise, and $\bm{e}_z$ the unit vector in $z$ direction. The decomposition of higher order moments into variance are due to the Wick theorem~\cite{peskin2018introduction} for Gaussian noise.

By recursively expanding the velocity Langevin equation one can show that the Wick theorem leads to exponential decay of the velocity autocorrelations; see \ref{sec:appendix_b} for this derivation.
Using that result, the MSD for \revise{non-interacting} ABPs is~\revise{\cite{romanczuk2012active,mikhailov1997self}}
\begin{equation}
\left< r^2 (t) \right> = 
v_0^2 \left< \int_0^t d\tau \int_0^t d\tau' e^{-\frac{\sigma^2}{2} \vert \tau-\tau' \vert} \right> 
=  \frac{4 v_0^2}{\sigma^2} \left(t - \frac{2}{\sigma^2} (1 - e^{-\frac{\sigma^2}{2} t})
\right),
\end{equation}
\revise{
which at long times satisfies the scaling relation
\begin{equation}
\left< r^2 (t) \right> \propto t \quad\mbox{as}\quad t \to \infty,
\label{eq:MSD-abp}
\end{equation}
while the MSD for non-interacting L\'evy swimmers scales as~\cite{zaburdaev2015levy}
\begin{equation}
\left< r^2 (t) \right> \propto t^{3-\alpha}
\quad\mbox{as}\quad t \to \infty.
\label{eq:MSD-levy}
\end{equation}
}
This reveals that ABPs exhibit behavior that is ballistic at short time scales, and normally diffusive at long time scales. This also holds for normal RTPs, as can be shown using Hilbert expansion analysis~\cite{othmer2000diffusion,plaza2019derivation}. 
\revise{In contrast, the L\'evy swimmers exhibit superdiffusive propagation at long times. The L\'evy swimmers also exhibit ballistic behavior at short time scales, but the ballistic front of the densities of ABP/RTP and L\'evy swimmers are qualitatively different. The short persistent times of ABP/RTP yields ballistic fronts that decay rapidly, leading to a self-similar property of the density in time characterized by a length scale proportional to $t^{1/2}$. This is consistent with the MSD scaling given by equation~\eqref{eq:MSD-abp}. In contrast, the ballistic front of the density of L\'evy swimmers, while decaying, persists to an extent that the density is not self-similar in time. However, the center of the density profile follows a self-similar scaling proportional to $t^{1/\alpha}$~\cite{zaburdaev2015levy}, which is distinct from the MSD scaling given by equation~\eqref{eq:MSD-levy}.}

Active L\'{e}vy swimmers share the discrete tumbling behavior as normal RTPs; neither involve translational diffusion. 
Our analysis and simulations show clearly a singular accumulation of active L\'evy swimmers on the walls of the channel, similar to RTPs. However,
most importantly, the correlation of L\'{e}vy swimmers in the bulk is qualitatively non-local and the density distributions converge to a non-uniform steady-state distribution inside the bulk at small $\ell/H$, which is well approximated by the steady-state solution \eqref{eqn:steady_state}. This novel aspect distinguishes active L\'evy swimmers from both RTPs and ABPs.

\section{Conclusion and Discussions}

We introduce a model for active swimmers with L\'{e}vy statistics under confinement. The L\'{e}vy swimmers are characterized by the power-law tail of their run-time distribution $\psi(\tau)$, which is taken as the Pareto distribution, and a constant swim force and hence a constant velocity magnitude $U_0$ between tumbling events when not interacting with other objects. On the domain boundaries we assume the L\'{e}vy swimmers follow the Skorokhod sliding condition, similar to the ABP and RTP models. Our modification to previous multiscale analysis (Hilbert expansion) shows that a fractional diffusion equation still holds as the leading order description in the weak confinement regime, with the Riesz fractional derivative and no-flux boundary conditions. In this asymptotic limit of $\ell/H \rightarrow 0$ the L\'{e}vy swimmers show qualitative difference from the ABP and RTP models, with a non-uniform U-shaped bulk density distribution. This is distinctive from the accumulation boundary layer effect of ABPs and RTPs, which scales with activity and dimishes as $\ell/H$ decreases, eventually recovering a uniform bulk distribution.
In the strong confinement limit \change{neither the L\'evy swimmers nor the RTPs have time} to tumble in the bulk of the channel, which results in a uniform bulk density profile. The fraction of swimmers accumulated and the force exerted on the walls are shown to converge to an analytic asymptote. Our analysis agree with particle simulations of stochastic trajectories of the swimmers.
Here, we show results for 1D and 2D only, but the  analysis can be easily generalized to 3D.

One advantage of our L\'{e}vy swimmer model compared to L\'{e}vy flights or walks is that the only prescription of swim force and drag coefficient readily admits extensions to interacting swimmers with finite sizes. In our preliminary simulations
we have observed motility-induced phase separation of L\'{e}vy swimmers interacting with the Weeks-Chandler-Andersen (WCA) potential~\cite{weeks1971role} at high packing density and activity. In future work, we will investigate how the phase diagram and the universality class may change from ABPs and normal RTPs to L\'{e}vy swimmers. \revise{It is also interesting to compare the behaviors of L\'evy swimmers with ABP/RTP in an external field.}

Our study of active L\'evy swimmers connects the physics of L\'evy walks with recent developments in fractional calculus. This provides a mathematical foundation to study L\'evy walks with interactions confined in more general geometries.
The distinguishing properties of active L\'evy swimmers, as compared to RTP and ABP models, show that they are a promising model for active matter in upstream swimming and transport in porous media. 

\ack
We thank Z. He, S. Yip, H. Row, Z.G. Wang and C. Kjeldbjerg for insightful discussions. T.Z. is supported by the Cecil and Sally Drinkward Postdoc Fellowship. T.Z. and M.G. are grateful for the opportunity to hold discussions at the 2018 ICERM workshop ``Fractional PDEs: Theory, Algorithms and Applications'' at Brown University. J.F.B. is supported by NSF grant CBET 1803662.

M.G. is supported by the John von Neumann fellowship at Sandia National Laboratories. Sandia National Laboratories is a multimission laboratory managed and operated by National Technology and Engineering Solutions of Sandia, LLC., a wholly owned subsidiary of  Honeywell  International,  Inc.,  for  the  U.S.  Department  of  Energy’s  National  Nuclear Security Administration under contract {DE-NA0003525}.  The views expressed in the article do not necessarily represent the views of the U.S. Department of Energy or the United States Government.  SAND number:  {SAND2021-2756 O}.

\appendix

\section{Details of particle simulations}\label{sec:simulation_details}
Our simulation data for Figures \ref{fig:1D-density}, \ref{fig:2D-density}, and \ref{fig:force-polarorder} were obtained from the stochastic dynamics of the L\'{e}vy swimmers described in Section \ref{sec:model}.  At the beginning of each simulation, 100,000 swimmers are released from randomized initial positions within the channel, with uniformly random initial orientations. A run-time sampled from the \change{type-II Pareto distribution $\psi(\tau)$ given by equation \eqref{eqn:lomax}} is assigned to each swimmer. After a swimmer runs for this time, it is assigned a new orientation from a uniform angular distribution. In 1D, this assigns probabilities of $1/2$ for both the left and right directions. In 2D, $\theta$ is distributed in $[0,2\pi)$. At the same time, a new run-time is sampled again from $\psi(\tau)$ and assigned to this swimmer. 
Between these tumbling events, a swimmer's trajectory is explicitly integrated with a timestep of $\Delta t=10^{-4}\tau_m$. When it hits the walls, the sliding no-flux boundary condition is imposed via a potential-free algorithm \cite{heyes1993brownian}.
In 2D, a periodic boundary condition is imposed for the direction along the channel. Each simulation is run for $O(10^2 \tau_m)$ to ensure sufficient convergence to the steady state, and statistics are measured by averaging over all swimmers for the last $O(10\tau_m)$ time interval.

\section{Kinetic equations at steady state}\label{sec:appendix_a}
At steady state, explicit dependence on time and on the direction $x$ parallel to the channel disappears.
Hence, in 2D the microscopic equations \eqref{eqn:micro-bulk}  -- \eqref{eqn:micro-flux-out-wall} reduce to 
\begin{equation}
\left\{
\begin{split}
\left(\frac{\partial}{\partial \tau} + U_0\sin\theta\frac{\partial}{\partial y}\right)P(y,\theta,\tau)& =-\beta(\tau)P(y,\theta,\tau)\\
\frac{\partial \phi^+}{\partial \tau}  & = -\beta(\tau)\phi^+ + j_b^+\\
j_b^+ &= U_0P(y=H/2,\tau,\theta^+)\sin\theta^+ \\
\phi^+(\tau=0,\theta^+) &= -P(y=H/2,\tau=0,\theta^-)U_0\sin\theta^- \\
&= \int_0^t \beta(\tau)d\tau \frac{1}{2\pi}\int_0^\pi d\theta' \phi^+ \\
&\approx \int_0^\infty \beta(\tau)d\tau \frac{1}{2\pi}\int_0^\pi d\theta' \phi^+.
\end{split}
\right.
\end{equation}
In 1D, the steady state equations are
\begin{equation}
\left\{
\begin{split}
\left(\frac{\partial}{\partial \tau}+U_0\frac{\partial}{\partial x}\right)P_{\pm}(x,\tau) & = -\beta(\tau)P_{\pm} \\
\int_0^{t\rightarrow\infty}\beta(\tau)d\tau P_{\pm} &= P_{\mp}(x,\tau=0)\\
\frac{\partial \phi^+}{\partial \tau} &= -\beta(\tau)\phi^+ + j_b^+\\
j_b^+(\tau) &= U_0 P_+(x=H/2,\tau) \\
\int_0^\infty d\tau\beta(\tau) \phi^+ &= U_0P_-(x=H/2,\tau=0).
\end{split}
\right.
\end{equation}

\section{Derivation of velocity autocorrelation function for ABPs}\label{sec:appendix_b}
Using a triple product identity and the fact that the particle moves in the $xy$ plane,
\begin{equation}
\bm{e}_z \times (\bm{e}_z \times \bm{v}) = (\bm{e}_z \cdot \bm{v}) \bm{e}_z - (\bm{e}_z \cdot \bm{e}_z) \bm{v} = -\bm{v}.
\end{equation}
Therefore, we can recursively expand the velocity as 
\begin{equation}
\begin{split}
\bm{e}_z \times \bm{v}(t) &= \bm{e}_z \times \left( \bm{v}_0 + \int_0^t d\tau\,  \eta(\tau) \bm{e}_z \times \bm{v}(\tau) \right) \\
&=  \bm{e}_z \times \bm{v}_0 - \int_0^t d\tau\, \eta(\tau) \bm{v}(\tau), \\
\end{split}
\end{equation}
where $\bm{v}_0$ is the initial velocity.
Hence 
\begin{equation}
\begin{split}
\left< \bm{v} (t) \right> 
& = \bm{v}_0 + \int_0^t dt_1 \left< \eta(t_1) \bm{e}_z \times \bm{v}(t_1) \right> \\
& = \bm{v}_0 + \int_0^t dt_1
\left(\left< \eta(t_1) \right> \bm{e}_z \times \bm{v}_0 - \int_0^{t_1} dt_2 \left< \eta(t_1)\eta(t_2) \bm{v}(t_2) \right>
\right) \\
& = \bm{v}_0 - \int_0^t dt_1
\int_0^{t_1} dt_2 \left< \eta(t_1)\eta(t_2) \bm{v}(t_2) \right> \\
& = \bm{v}_0 - \int_0^t dt_1
\int_0^{t_1} dt_2 \left< \eta(t_1)\eta(t_2) \right>
 + \int_0^t 
\int_0^{t_1}  \int_0^{t_2} \int_0^{t_3} \prod_{i=1}^4 dt_i \left< \eta(t_1)\eta(t_2)  \eta(t_3)\eta(t_4) \right> - ... \\
& = \bm{v}_0 \left( 1 - t \sigma^2/2 + \frac{t^2}{2} \sigma^4/4\right) - \int...\left<...\bm{v}(t_6)\right>\\
& = \bm{v}_0 \sum_{n=0}^\infty \frac{1}{n!} (-t\sigma^2/2)^n \\
& = \bm{v}_0 e^{-t\sigma^2/2}.
\end{split}
\end{equation}
The above derivation follows from applying the Wick theorem~\cite{peskin2018introduction}. Due to the hierarchy of upper limits of integrations, only the leading term of the Wick summation survives. The evaluation of the integrals uses equation \eqref{eqn:eta_definition} and the fact that
\begin{equation}
\delta(x-y) = \lim_{\tau\rightarrow{0}} \begin{cases}
1/\tau & \mbox{ if } \vert x-y \vert < \tau \\
0 & \mbox{ otherwise}.
\end{cases}
\end{equation}

\newcommand{\newblock}{}
\bibliographystyle{unsrtnat}
\bibliography{als}

\end{document}